\documentclass[12pt]{article}
\usepackage{amsfonts,amssymb,amsmath,mathrsfs}
\usepackage{hyperref}
\newcommand{\beq}{\begin{equation}}
\newcommand{\eeq}{\end{equation}}
\newcommand{\vp}{\vphantom}
\begin{document}
\begin{center}
{\Large\bf Lagrangian mechanics of massless superparticle\\[0.3cm] on $AdS_4\times\mathbb{CP}^3$ superbackground}\\[0.5cm]
{\large D.V.~Uvarov\footnote{E-mail: d\_uvarov@\,hotmail.com}}\\[0.2cm]
{\it NSC Kharkov Institute of Physics and Technology,}\\ {\it 61108 Kharkov, Ukraine}\\[0.5cm]
\end{center}
\begin{abstract}
Massless superparticle model is considered on the $OSp(4|6)/(SO(1,3)\times U(3))$ supercoset manifold and in the $AdS_4\times\mathbb{CP}^3$ superspace. In the former case integrability of the equations of motion is rather obvious, while for the $AdS_4\times\mathbb{CP}^3$ superparticle we prove integrability in the partial $\kappa-$symmetry gauge for which 4 anticommuting coordinates related to the broken conformal supersymmetry are set to zero. This allows us to propose expression for the Lax pair that may encode complete equations of motion for the $AdS_4\times\mathbb{CP}^3$ superparticle.
\end{abstract}

\section{Introduction}

Aharony-Bergman-Jafferis-Maldacena (ABJM) correspondence \cite{ABJM} identifying $D=3$ $\mathcal N=6$ superconformal Chern-Simons-matter theory with $U(N)_k\times U(N)_{-k}$ gauge symmetry and M-theory on $AdS_4\times(S^7/\mathbb Z_k)$ background belongs to the family of $AdS/CFT$ dualities \cite{Maldacena}. The distinctive feature of ABJM correspondence as compared to the $AdS_5/CFT_4$ one -- the most notable member of the family -- is non-maximal supersymmetry of the dual theories. So the $AdS_4\times(S^7/\mathbb Z_k)$ background of $11d$ supergravity preserves 24 of 32 supersymmetries, the same amount of supersymmetry has the $AdS_4\times\mathbb{CP}^3$ background of $D=10$ IIA supergravity \cite{Watamura}, \cite{NPope}, \cite{STV} that arises in the limit $k^5\gg N\gg1$. This fact limits straight-forward generalization of the results obtained for the $AdS_5/CFT_4$ correspondence to the ABJM duality case and issues new challenges.

Even the construction of the IIA superstring action on
$AdS_4\times\mathbb{CP}^3$ superbackground required non-trivial
efforts. The group-theoretic supercoset approach \cite{MT98},
initially elaborated to construct the IIB superstring on
$AdS_5\times S^5$ superspace, gives only part of the
$AdS_4\times\mathbb{CP}^3$ superstring action that is described by
the $OSp(4|6)/(SO(1,3)\times U(3))$ sigma-model \cite{AF},
\cite{Stefanski}\footnote{Alternative approach to constructing the
$OSp(4|6)/(SO(1,3)\times U(3))$ sigma-model action uses pure
spinor variables \cite{PS}.}. Application of the supercoset
approach is based on the isomorphism between the
$OSp(4|6)/(SO(1,3)\times U(3))$ supercoset manifold and the
subspace of $AdS_4\times\mathbb{CP}^3$ superspace parametrized by
10 space-time and 24 anticommuting coordinates associated with the
unbroken supersymmetries of the background. The full-fledged
$AdS_4\times\mathbb{CP}^3$ superspace parametrized additionally by
8 Grassmann coordinates related to the broken space-time
supersymmetries cannot be described as a supercoset manifold. A
way to recover its supergeometry is the reduction of that for the
maximally supersymmetric $AdS_4\times S^7$ superspace, isomorphic
to the $OSp(4|8)/(SO(1,3)\times SO(7))$ supercoset manifold, using
the Hopf fibration realization of the 7-sphere
$S^7=\mathbb{CP}^3\times S^1$ \cite{NPope}, \cite{STV}.
Accordingly it was suggested in \cite{AF} that the complete
$AdS_4\times\mathbb{CP}^3$ superstring action can be obtained by
the double-dimensional reduction \cite{DHIS}, \cite{HoweSezgin} of
the $AdS_4\times S^7$ supermembrane \cite{de Wit}. Geometric
constituents of the $AdS_4\times\mathbb{CP}^3$ superspace were
found and applied to the construction of complete superstring
action in \cite{GSWnew}\footnote{The $OSp(4|6)/(SO(1,3)\times
U(3))$ sigma-model action comes about upon partial $\kappa-$symmetry gauge
fixing of the $AdS_4\times\mathbb{CP}^3$ superstring by setting to
zero 8 Grassmann coordinates related to broken supersymmetries.
This constrains the range of application of the
$OSp(4|6)/(SO(1,3)\times U(3))$ sigma-model \cite{AF},
\cite{GSWnew}.}. Its
structure, however, appears to be complicated highly non-linear.
Attempts \cite{GrSW}, \cite{U09} to simplify it by appropriate
choice of gauges for local symmetries, in particular, the
$\kappa-$symmetry have not resulted so far in a simple enough
gauge-fixed action. The similar situation occurred with the
$AdS_5\times S^5$ superstring \cite{Pesando},
\cite{KalloshRahmfeld}, \cite{KT98}, \cite{MT2000}.

The renewed impetus to exploring the $AdS_5/CFT_4$ correspondence
was the discovery \cite{BPR} of the integrability of $AdS_5\times
S^5$ superstring equations of motion supported by unveiling
integrable structure in the dual $\mathcal N=4$ super-Yang-Mills
theory \cite{Minahan} (see \cite{Beisert} for a recent review).
The integrability of equations of motion for the
$OSp(4|6)/(SO(1,3)\times U(3))$ sigma-model \cite{AF},
\cite{Stefanski} is the straight-forward generalization of that for the
$AdS_5\times S^5$ superstring\footnote{In fact both models belong
to the class of integrable sigma-models on supercoset manifolds with
the $\mathbb Z_4-$graded isometry superalgebras \cite{Adam},
\cite{BSZ}, \cite{STWZ}.}. This allowed to work out corresponding
algebraic curve \cite{0807.0437} and all-loop Bethe equations
\cite{0807.0777}, \cite{Ahn}.

Non-trivial problem is an extension of the integrable structure to the complete $AdS_4\times\mathbb{CP}^3$ superstring case. Evidence supporting the possibility of such an extension was given in \cite{SW10}, \cite{CSW}. Since the Lax connection encoding the $OSp(4|6)/(SO(1,3)\times U(3))$ sigma-model equations takes value in the $osp(4|6)$ isometry superalgebra and is presented as the linear combination of $osp(4|6)$ Cartan forms it is reasonable to suggest that the Lax connection for the $AdS_4\times\mathbb{CP}^3$ superstring is a generalization of that for the sigma-model by the contributions of fermionic coordinates related to 8 broken supersymmetries and their differentials. The authors of \cite{CSW} succeeded in extending the $OSp(4|6)/(SO(1,3)\times U(3))$ sigma-model Lax connection by linear and quadratic terms in those 'broken' fermions with its curvature turning to zero on the $AdS_4\times\mathbb{CP}^3$ superstring equations up to quadratic order also. Obtained expression suggests that complete Lax connection for the  $AdS_4\times\mathbb{CP}^3$ superstring has a rather complicated form.

In \cite{U12} based on the isomorphism between the $osp(4|8)$ superalgebra and $D=3$ $\mathcal N=8$ superconformal algebra we proposed to use the $\kappa-$symmetry gauge freedom to single out 4 minimal $\frac14$ fractions of broken supersymmetries described by $SL(2,\mathbb R)$ (anti)Majorana fermions associated with the generators of broken Poincare and conformal supersymmetries.\footnote{Coordinates associated with the superconformal algebra generators were introduced in  \cite{9807115}, \cite{Kallosh2}, \cite{PST}, \cite{MT2000} to describe the string/brane models related to the maximally supersymmetric examples of $AdS/CFT$ correspondence.} This allows to introduce a minimal extension of the $OSp(4|6)/(SO(1,3)\times U(3))$ sigma-model. In the case when in the sector of broken supersymmetries there have been gauged away all the coordinates except for the $SL(2,\mathbb R)$ Majorana spinor related to the broken Poincare supersymmetries\footnote{Allowing fermionic coordinates related to broken supersymmetries to be among the physical modes relaxes restrictions on the superstring motions that arise for the $OSp(4|6)/(SO(1,3)\times U(3))$ sigma-model. For instance the string configurations lying within the $AdS_4$ part of the background that cannot be described by the $OSp(4|6)/(SO(1,3)\times U(3))$ sigma-model can be considered in such a gauge.} we have shown that the corresponding extension of the $OSp(4|6)/(SO(1,3)\times U(3))$ sigma-model is classically integrable with the Lax connection whose curvature strictly equals zero and that coincides with the subsector of the Lax connection found in \cite{CSW}. Considering other $\kappa-$gauges that leave in the broken supersymmetries sector two of more $\frac14$ fractions it might be possible to extend the $OSp(4|6)/(SO(1,3)\times U(3))$ sigma-model Lax connection by corresponding $SL(2,\mathbb R)$ (anti)Majorana coordinates. In particular, there are 6 ways to make up $\frac12$ fraction of broken supersymmetries. One can choose anticommuting coordinates related to Poincare and conformal supersymmetries that satisfy (anti)Majorana condition. These two options are covered by the Kaluza-Klein gauge condition \cite{U12} characterized by vanishing of those summands in the $AdS_4\times S^7$ supervielbein bosonic components in directions tangent to anti-de Sitter space-time that are proportional to the differential of the $S^1$ fiber coordinate $dy$. Two other possibilities correspond to choosing Majorana coordinate for the broken Poincare supersymmetries and anti-Majorana coordinate for the broken conformal supersymmetries and vice versa. Finally one can consider unconstrained $D=3$ fermion $\theta^\mu$ and its conjugate $\bar\theta^\mu$ associated with the generators of broken Poincare supersymmetries or unconstrained spinors $\eta_\mu$ and $\bar\eta_\mu$ related to the generators of broken conformal supersymmetries.\footnote{These two options correspond to the $\kappa-$symmetry gauge conditions of Ref.~\cite{GrSW} restricted to the 'broken' fermions.} In these four latter cases the $AdS_4\times S^7$ supervielbein bosonic components tangent to $AdS_4$ do have non-zero summands proportional to $dy$ so that the tangent-space Lorentz rotation needs to be performed to remove them in accordance with the general prescription of the $D=10$ supervielbein construction \cite{DHIS}, \cite{HoweSezgin}.

So as a further step towards recovering the
$AdS_4\times\mathbb{CP}^3$ superstring integrable structure one
can consider the extension of the $OSp(4|6)/(SO(1,3)\times U(3))$
sigma-model by the $\theta^\mu$, $\bar\theta^\mu$ Grassmann
coordinates associated with the broken Poincare supersymmetries.
This case combines reasonable simplicity with the possibility to
probe those terms in the $AdS_4\times\mathbb{CP}^3$ supervielbein
that arise as a result of the Lorentz rotation. To simplify the
matter as much as possible we concentrate on the zero modes sector described by the massless
superparticle model.\footnote{It can be viewed as the mass-to-zero
limit of the D0-brane on $AdS_4\times\mathbb{CP}^3$
superbackground \cite{GrSW}.} As a warm up we start with the
superparticle on the $OSp(4|6)/(SO(1,3)\times U(3))$ supercoset
space introduced in Ref.~\cite{Stefanski}, derive the equations of
motion and find their Lax representation. Then we examine the case
of superparticle on the $AdS_4\times\mathbb{CP}^3$ superbackground
with $\eta_\mu$, $\bar\eta_\mu$ coordinates gauged away and obtain
the Lax representation for the equations of motion in such a partial
$\kappa-$symmetry gauge. This allows to make a prediction for the
form of the Lax pair of the superparticle in complete $AdS_4\times\mathbb{CP}^3$
superspace. We finish by discussing what can be learnt on the
$AdS_4\times\mathbb{CP}^3$ superstring Lax connection from the
superparticle Lax pair.

\section{Massless superparticle on the $OSp(4|6)/(SO(1,3)\times U(3))$ supermanifold}
\subsection{$osp(4|6)$ superalgebra, Cartan forms and $\mathbb Z_4-$grading}

Taking an $OSp(4|6)/(SO(1,3)\times U(3))$ representative $\mathscr G$, left-invariant $osp(4|6)$ Cartan forms in conformal basis \cite{U08} can be grouped as
\beq
\label{defcartan}
\mathscr C(d)=\mathscr G^{-1}d\mathscr G=\mathcal C_{\mbox{\scriptsize conf}}+\mathcal C_{\mbox{\scriptsize su(4)}}+\mathcal C_{\mbox{\scriptsize 24susys}}.
\eeq
The first summand introduces Cartan forms associated with the $D=3$ conformal group generators $(D, P_m, K_m, G_{mn})$
\beq\label{defcartan1}
\mathcal C_{\mbox{\scriptsize conf}}=\Delta(d)D+\omega^m(d)P_m+c^m(d)K_m
+G^{mn}(d)M_{mn}.
\eeq
Other bosonic Cartan forms are associated with the $so(6)\sim su(4)$ $R-$symmetry group generators
\beq\label{defcartan2}
\mathcal C_{\mbox{\scriptsize su(4)}}=\Omega_a(d)T^a
+\Omega^a(d)T_a+\widetilde\Omega_a{}^b(d)\widetilde V_b{}^a+\widetilde\Omega_a{}^a(d)\widetilde V_b{}^b.
\eeq
They have been divided into $\widetilde V_b{}^a$ generators of the $U(3)$ stability group of $\mathbb{CP}^3=SU(4)/U(3)$ manifold and the $su(4)/u(3)$ coset generators $(T_a, T^a)$.
The last summand in (\ref{defcartan}) includes odd Cartan forms
\beq\label{defcartan3}
\mathcal C_{\mbox{\scriptsize 24susys}}=\omega^\mu_a(d)Q^a_\mu+\bar\omega^{\mu a}(d)\bar Q_{\mu a}+\chi_{\mu
a}(d)S^{\mu a}+\bar\chi^a_\mu(d)\bar S^\mu_a,
\eeq
where the generators of $D=3$ $\mathcal N=6$ super-Poincare $(Q^a_\mu, \bar Q_{\mu a})$ and conformal $(S^{\mu a}, \bar S^\mu_a)$ supersymmetries carry $SL(2,\mathbb{R})$ spinor index $\mu=1,2$ and $SU(3)$ (anti)fundamental representation index $a=1,2,3$ in accordance with the decomposition of $SO(6)$ vector on the $SU(3)$ representations $\mathbf 6=\mathbf 3\oplus\bar{\mathbf 3}$.

(Anti)commutation relations of the $osp(4|6)$ superalgebra are known to be invariant under 4-element discrete automorphism $\Upsilon$ that assigns different eigenvalues to the $osp(4|6)$ generators
\beq
\Upsilon(g_{(k)})=i^kg_{(k)},\ k=0,1,2,3:\quad[g_{(j)},g_{(k)}\}=g_{(j+k)mod4}.
\eeq
So that (\ref{defcartan}) can be written in the $\mathbb{Z}_4-$graded form
\beq\label{gradedcf}
\mathscr C(d)=\mathscr C_{(0)}(d)+\mathscr C_{(2)}(d)+\mathscr C_{(1)}(d)+\mathscr C_{(3)}(d),
\eeq
where
\beq
\begin{array}{rl}
\mathscr C_{(0)}(d)=&2G^{3m}(d)M_{3m}+G^{mn}(d)M_{mn}+\widetilde\Omega_a{}^b(d)\widetilde V_b{}^a+\widetilde\Omega_a{}^a(d)\widetilde V_b{}^b\in g_{(0)},\\[0.2cm]
\mathscr C_{(2)}(d)=&2G^{0'm}(d)M_{0'm}+\Delta(d)D+\Omega_a(d)T^a+\Omega^a(d)T_a\in g_{(2)},\\[0.2cm]
\mathscr C_{(1)}(d)=&\omega_{(1)}\vp{\omega}^\mu_{a}(d)Q_{(1)}\vp{Q}^a_\mu+\bar\omega^{\vp{\mu c}}_{(1)}\vp{\bar\omega}^{\mu a}(d)\bar Q_{(1)\mu a}\in g_{(1)},\\[0.2cm]
\mathscr C_{(3)}(d)=&\omega_{(3)}\vp{\omega}^\mu_{a}(d)Q_{(3)}\vp{Q}^a_\mu+\bar\omega^{\vp{\mu c}}_{(3)}\vp{\bar\omega}^{\mu a}(d)\bar Q_{(3)\mu a}\in g_{(3)}.
\end{array}
\eeq
In the above formulae we introduced the $so(2,3)$ algebra generators $M_{0'm}=\frac12(P_m+K_m)$, $M_{0'3}=-D$, $M_{3m}=\frac12(K_m-P_m)$ and fermionic generators with definite $\mathbb Z_4-$grading
\beq\label{gradedfermidef}
Q^{\vp{a}}_{(1)}\vp{Q}^a_\mu=Q^a_\mu+iS^a_\mu,\quad\bar Q_{(1)}\vp{Q}_{\mu a}=\bar Q_{\mu a}-i\bar S_{\mu a};\quad Q^{\vp{a}}_{(3)}\vp{Q}^a_\mu=Q^a_\mu-iS^a_\mu,\quad\bar Q_{(3)}\vp{Q}_{\mu a}=\bar Q_{\mu a}+i\bar S_{\mu a}.
\eeq
Associated bosonic
\beq
G^{0'm}(d)=\frac12(\omega^m(d)+c^m(d)),\quad G^{3m}(d)=-\frac12(\omega^m(d)-c^m(d))
\eeq
and fermionic Cartan forms
\beq
\omega_{(1)}\vp{\omega}^\mu_{a}(d)=\frac12(\omega^\mu_a(d)+i\chi^\mu_a(d)),\quad\omega_{(3)}\vp{\omega}^\mu_{a}(d)=\frac12(\omega^\mu_a(d)-i\chi^\mu_a(d))
\eeq
and c.c. are used to work out the Lax representation for superparticle equations of motion. Let us note that Cartan forms from $\mathscr C_{(2)}$ and $\mathscr C_{(1,3)}$ eigenspaces are identified with the bosonic and fermionic components of $OSp(4|6)/(SO(1,3)\times U(3))$ supervielbein, while those from $\mathscr C_{(0)}$ eigenspace describe $SO(1,3)\times U(3)$ connection.

\subsection{$OSp(4|6)/(SO(1,3)\times U(3))$ superparticle action and equations of motion}

Massless superparticle action on the $OSp(4|6)/(SO(1,3)\times U(3))$ supercoset manifold \cite{Stefanski}
\beq\label{cosetaction}
\mathscr S_{\mbox{\scriptsize{coset}}}=\int\frac{d\tau}{e}(G_{\tau}\vp{G}^{0'}\vp{G}_{m}G_\tau\vp{G}^{0'm}+\Delta_\tau\Delta_\tau+\Omega_{\tau a}\Omega_\tau{}^a),
\eeq
constructed out of the world-line pullbacks of Cartan forms from the $\mathscr C_{(2)}$ eigenspace, is invariant under the $\mathbb{Z}_4$ automorphism and $OSp(4|6)$ global symmetry acting as the left group multiplication on $\mathscr G$. Action variation w.r.t. the Lagrange multiplier $e(\tau)$ produces the mass-shell constraint
\beq
G_{\tau}\vp{G}^{0'}\vp{G}_{m}G_\tau\vp{G}^{0'm}+\Delta_\tau\Delta_\tau+\Omega_{\tau a}\Omega_\tau{}^a=0.
\eeq
Since it is irrelevant to the subsequent discussion of the Lax representation for other (dynamical) equations of motion we set the Lagrange multiplier to unity. In this respect it can be said that we consider a '$1d$ sigma model'.

To derive dynamical equations from (\ref{cosetaction}) it is convenient to use the following general relation for the variation of a 1-form
\beq
\delta G(d)=di_\delta G+i_\delta dG
\eeq
and substitute in the second summand Maurer-Cartan equations for $\mathcal C_{(2)}$ Cartan forms \cite{U08}
\beq
\begin{array}{rl}
dG^{0'm}-&2G^{3m}(d)\wedge\Delta(d)-2G^{mn}(d)\wedge G^{0'}\vp{G}_n(d)+2i\omega_{(1)}\vp{\omega}^\mu_{a}(d)\wedge\sigma^m_{\mu\nu}\bar\omega_{(1)}\vp{\bar\omega}^{\nu a}(d)\\[0.2cm]
+&2i\omega_{(3)}\vp{\omega}^\mu_{a}(d)\wedge\sigma^m_{\mu\nu}\bar\omega_{(3)}\vp{\bar\omega}^{\nu a}(d)=0,\\[0.2cm]
d\Delta+&2G^{3m}(d)\wedge G^{0'}\vp{G}_m(d)+2\omega_{(1)}\vp{\omega}^\mu_{a}(d)\wedge\bar\omega_{(1)}\vp{\bar\omega}_\mu^{a}(d)-2\omega_{(3)}\vp{\omega}^\mu_{a}(d)\wedge\bar\omega_{(3)}\vp{\bar\omega}_\mu^{a}(d)=0,\\[0.2cm]
d\Omega^a+&i\Omega^b(d)\wedge(\widetilde\Omega_b{}^a(d)+\delta^a_b\widetilde\Omega_c{}^c(d))-2i\varepsilon^{abc}\omega_{(1)}\vp{\omega}^\mu_{b}(d)\wedge\omega_{(1)\mu c}(d)\\[0.2cm]
+&2i\varepsilon^{abc}\omega_{(3)}\vp{\omega}^\mu_{b}(d)\wedge\omega_{(3)\mu c}(d)=0.
\end{array}
\eeq
Then taking the contractions of the Cartan forms (\ref{gradedcf}) with the variation symbol $i_\delta$ as independent parameters yields the set of equations of motion
\beq\label{coseteom}
\begin{array}{rl}
-\frac12\frac{\delta\mathscr S_{\mbox{\scriptsize{coset}}}}{\delta G^{0'}{}_m(\delta)}=&\frac{dG_\tau\vp{G}^{0'm}}{d\tau}+2G_\tau\vp{G}^{mn}G_\tau\vp{G}^{0'}\vp{G}_{n}+2G_\tau\vp{G}^{3m}\Delta_\tau=0,\\[0.2cm]
-\frac12\frac{\delta\mathscr S_{\mbox{\scriptsize{coset}}}}{\delta\Delta(\delta)}=&\frac{d\Delta_\tau}{d\tau}-2G_\tau\vp{G}^{3m}G_\tau\vp{G}^{0'}\vp{G}_m=0,\\[0.2cm]
-\frac{\delta\mathscr S_{\mbox{\scriptsize{coset}}}}{\delta\Omega_a(\delta)}=&\frac{d\Omega_\tau\vp{\Omega}^a}{d\tau}+i\Omega_\tau\vp{\Omega}^b(\widetilde\Omega_{\tau b}{}^a+\delta_b^a\widetilde\Omega_{\tau c}{}^c)=0;\\[0.2cm]
-\frac14\frac{\delta\mathscr S_{\mbox{\scriptsize{coset}}}}{\delta\bar\omega^{\vp{a}}_{(1)}\vp{\bar\omega}^{a}_\mu(\delta)}=&iG_\tau\vp{G}^{0'm}\tilde\sigma_{m}^{\mu\nu}\omega_{(1)\tau\nu a}+\Delta_\tau\omega_{(1)}\vp{\omega}^{\vp{\mu}}_{\tau}\vp{\omega}^\mu_{a}-i\varepsilon_{abc}\Omega_\tau{}^b\bar\omega^{\vp{\mu c}}_{(1)\tau}\vp{\bar\omega}^{\mu c}=0;\\[0.2cm]
-\frac14\frac{\delta\mathscr S_{\mbox{\scriptsize{coset}}}}{\delta\bar\omega^{\vp{a}}_{(3)}\vp{\bar\omega}^{a}_\mu(\delta)}=&iG_\tau\vp{G}^{0'm}\tilde\sigma_{m}^{\mu\nu}\omega_{(3)\tau\nu a}-\Delta_\tau\omega_{(3)}\vp{\omega}^{\vp{\mu}}_{\tau}\vp{\omega}^\mu_{a}+i\varepsilon_{abc}\Omega_\tau{}^b\bar\omega^{\vp{\mu c}}_{(3)\tau}\vp{\bar\omega}^{\mu c}=0
\end{array}
\eeq
and c.c. equations\footnote{We assume right derivative for fermions.}.

\subsection{Lax representation for the $OSp(4|6)/(SO(1,3)\times U(3))$ superparticle equations of motion}

Eqs. (\ref{coseteom}) admit the Lax representation
\beq\label{cosetlax}
\frac{d\mathfrak l}{d\tau}+[M,\mathfrak l]=0
\eeq
with $M=\mathscr C_\tau$ determined by the pullback of Cartan forms (\ref{gradedcf}) and
\beq
\mathfrak l=2G_\tau\vp{G}^{0'm}M_{0'm}+\Delta_\tau D+\Omega_{\tau a}T^a+\Omega_\tau{}^aT_a\in g_{(2)}.
\eeq
Let us note the ambiguity in the definition of $M$. Instead of $\mathscr C_\tau$ it can be defined as $\mathscr C_{(0)\tau}\oplus\mathscr C_{(1)\tau}\oplus\mathscr C_{(3)\tau}$. This ambiguity is resolved when the 'broken' Grassmann coordinates are taken into consideration in favor of the first possibility, i.e. $M=\mathscr C_\tau$.

Since $\mathfrak l$ is determined by the Cartan forms that enter the superparticle action (\ref{cosetaction}) one can write it down in the form of a  differential operator from $g_{(2)}$ eigenspace acting on the action functional
\beq\label{diffoper}
\mathfrak l=\left(M_{0'm}\frac{\partial}{\partial G_{\tau}\vp{G}^{0'}\vp{G}_m}+\frac12D\frac{\partial}{\partial\Delta_{\tau}}+T^a\frac{\partial}{\partial\Omega_{\tau}\vp{\Omega}^a}+T_a\frac{\partial}{\partial\Omega_{\tau a}}\right)\mathscr S_{\mbox{\scriptsize{coset}}}.
\eeq
Below we shall observe how this Lax pair generalizes by inclusion of the 'broken' fermions.

When the $OSp(4|6)/(SO(1,3)\times U(3))$ superparticle action is considered as arising from that of the superparticle on $AdS_4\times\mathbb{CP}^3$ superbackground in the partial $\kappa-$symmetry gauge characterized by setting to zero 8 'broken' Grassmann coordinates there remain non-trivial equations for those coordinates
\beq\label{orthogeom}
\Omega_{\tau a}\bar\chi^{\vp{\mu}}_\tau\vp{\bar\chi}^{\mu a}=0,\quad\Omega_{\tau a}\bar\omega_\tau\vp{\bar\omega}^{\mu a}=0
\eeq
and c.c. They can be shown \cite{U-Tomsk} to follow from the fermionic equations of the $OSp(4|6)/(SO(1,3)\times U(3))$ superparticle so that all the independent equation of motion are given by the system (\ref{coseteom}).\footnote{Demonstrating that equations for the fermions associated with the broken supersymmetries constitute a part of those for the $OSp(4|6)/(SO(1,3)\times U(3))$ superparticle/sigma-model provides necessary consistency check of the corresponding $\kappa-$symmetry gauge condition. To leading order in the Grassmann coordinates of the $OSp(4|6)/(SO(1,3)\times U(3))$ supermanifold this was shown in \cite{SW10} and the general proof was given in \cite{U-Tomsk}.} This is not the case for other choices of the $\kappa-$gauge when the fermionic coordinates associated with the broken supersymmetries are among the physical degrees of freedom.

\section{Massless superparticle on the $AdS_4\times\mathbb{CP}^3$ superbackground}

\subsection{$AdS_4\times S^7$ superbackground, partial $\kappa-$symmetry gauge fixing and reduction to $D=10$}

Geometric constituents of the $AdS_4\times\mathbb{CP}^3$ superspace parametrized by all 32 Grassmann coordinates can be derived from those of the $AdS_4\times S^7$ superspace by means of the dimensional reduction \cite{GSWnew}, \cite{GrSW}. The maximally supersymmetric $AdS_4\times S^7$ superspace is isomorphic to the $OSp(4|8)/(SO(1,3)\times SO(7))$ supercoset manifold with the $OSp(4|8)$ isometry supergroup. Hence the $D=11$ supervielbein, connection and 4-form field strength are constructed out of the $osp(4|8)$ Cartan forms that can be presented in the $AdS_4\times S^7$ or supermembrane basis as
\begin{equation}\label{cf}
\hat{\mathscr G}^{-1}d\hat{\mathscr G}=2\mathcal G^{0'm'}(d)M_{0'm'}+\mathcal G^{m'n'}(d)M_{m'n'}+\Omega_{8I'}(d)V_{8I'}+\Omega_{I'J'}(d)V_{I'J'}+F^{\alpha
A'}(d)O_{\alpha A'},
\end{equation}
where $\hat{\mathscr G}$ is an $OSp(4|8)/(SO(1,3)\times SO(7))$ representative. $so(2,3)/so(1,3)\oplus so(8)/so(7)$ Cartan forms are identified with the
$AdS_4\times S^7$ supervielbein bosonic components
\beq\label{11dvielgen}
\hat E^{m'}(d)=\mathcal G^{0'm'}(d),\quad\hat E_{I'}(d)=\Omega_{8I'}(d),\quad m'=0,...,3;\ I'=1,...,7
\eeq
and fermionic Cartan forms -- with the supervielbein fermionic components
$F^{\alpha A'}(d)$
carrying indices $\alpha=1,...,4$ of the Majorana spinor of $Spin(1,3)$ and $A'=1,...,8$ of one of the $Spin(8)$ chiral spinor representations. Other Cartan forms describe the $so(1,3)\times so(7)$ connection. Field strength of the 3-form potential takes the form
\begin{equation}\label{h4}
H_{(4)}=\frac{i}{8}F^{\hat\alpha}\wedge\mathfrak{g}^{\hat m\hat
n}{}_{\hat\alpha}{}^{\hat\beta}F_{\hat\beta} \wedge\hat E_{\hat
m}\wedge\hat E_{\hat n}+\frac14\varepsilon_{m'n'k'l'}\hat E^{m'}\wedge\hat
E^{n'}\wedge\hat E^{k'}\wedge\hat E^{l'},
\end{equation}
where $\mathfrak{g}^{\hat m\hat
n}{}_{\hat\alpha}{}^{\hat\beta}$ stands for the antisymmetrized product of $D=11$ gamma-matrices $\mathfrak g^{\hat m}{}_{\hat\alpha}{}^{\hat\beta}$ with $\hat m=m'\oplus I'$ and $\hat\alpha=\alpha A'$, $\hat\beta=\beta B'$ being vector and spinor indices.\footnote{More details on the notation and spinor algebra can be found in \cite{U09}.} $D=11$ supervielbein bosonic components (\ref{11dvielgen}) and 4-form field strength (\ref{h4}) enter the $AdS_4\times S^7$ supermembrane action \cite{de Wit} that can be dimensionally reduced to the Type IIA superstring action on $AdS_4\times\mathbb{CP}^3$ superspace \cite{GSWnew}.

Dimensional reduction to the $AdS_4\times\mathbb{CP}^3$ superspace is based on the Hopf fibration realization of the 7-sphere $S^7=\mathbb{CP}^3\times S^1$ \cite{NPope}, \cite{STV}. It requires identifying mutually commuting $su(4)$ and $u(1)$ isometry algebras of $\mathbb{CP}^3$ and $S^1$ respectively within the $so(8)$ algebra of $S^7$. To this end the change of basis for the $so(8)$ generators $(V_{8I'}, V_{I'J'})$ needs to be performed \cite{GSWnew}, \cite{U09}. The $su(4)\sim so(6)$ becomes the bosonic subalgebra of $osp(4|6)$ isometry superalgebra of the $AdS_4\times\mathbb{CP}^3$ superbackground. 24 Fermionic generators of $osp(4|6)$ superalgebra are in one-to-one correspondence with the space-time supersymmetries of $AdS_4\times\mathbb{CP}^3$ superspace. They, as well as the generators of 8 supersymmetries broken by the $AdS_4\times\mathbb{CP}^3$ superbackground, can be extracted from 32  fermionic generators of $osp(4|8)$ superalgebra by acting with the projection matrices \cite{NPope}, \cite{GSWnew}, whose form is determined by the K\" ahler 2-form of $\mathbb{CP}^3$. Particularly convenient realization for that tensor, in which these projectors diagonalize, is in terms of $D=6$ chiral gamma-matrices \cite{U10}. Further using the isomorphism between the $osp(4|8)$ superalgebra and $D=3$ $\mathcal N=8$ superconformal algebra \cite{U09}, decomposing $D=6$ vectors and $D=8$ spinors under $SU(3)$ as $\mathbf 6=\mathbf 3\oplus\bar{\mathbf 3}$ and $\mathbf 8=\mathbf 4\oplus\bar{\mathbf 4}=(\mathbf 3\oplus\mathbf 1)\oplus(\bar{\mathbf 3}\oplus\bar{\mathbf 1})$ allows to present (\ref{cf}) as
\beq
%\begin{array}{rl}
\hat{\mathscr G}^{-1}d\hat{\mathscr G}=\mathcal C_{\mbox{\scriptsize\underline{conf}}}+\mathcal C_{\mbox{\scriptsize so(8)}}+\mathcal C_{\mbox{\scriptsize 32susys}}.
%\end{array}
\eeq
The first summand
\beq\label{confbasis1}
\mathcal C_{\mbox{\scriptsize\underline{conf}}}=\underline\Delta(d)D+
\underline{\omega}^m(d)P_m+\underline c^m(d)K_m
+\underline G^{mn}(d)M_{mn}
\eeq
contains Cartan forms related to $D=3$ conformal group generators similarly to (\ref{defcartan1}). In the above and subsequent relations we underline those $osp(4|6)$ Cartan forms that acquire dependence on 8 Grassmann coordinates associated with the broken supersymmetries and the $S^1$ coordinate differential  $dy$ in addition to the coordinates parametrizing the $OSp(4|6)/(SO(1,3)\times U(3))$ manifold. The second summand
\beq\label{hopfbasis}
\begin{array}{rl}
\mathcal C_{\mbox{\scriptsize so(8)}}=&\Omega_a(d)T^a+\Omega^a(d)T_a+\widetilde\Omega_a(d)\widetilde T^a+\widetilde\Omega^a(d)\widetilde T_a\\[0.2cm]
+&\widetilde\Omega_a{}^b(d)\widetilde V_b{}^a+\widetilde\Omega_b{}^b(d)\widetilde
V_a{}^a+\Omega_a{}^4(d)V_4{}^a+\Omega_4{}^a(d)V_a{}^4+h(d)H
\end{array}
\eeq
introduces Cartan forms for the $so(8)$ generators in the basis adapted for the Hopf fibration realization of the 7-sphere. The generators $(T_a, T^a, \widetilde V_b{}^a)$ form the $su(4)\subset so(8)$ algebra (cf. (\ref{defcartan2})). The generator $H$ corresponds to the $u(1)$ isometry of $S^1$ commuting with $su(4)$, while remaining 12 generators parametrize the $so(8)/(su(4)\times u(1))$ coset. Explicit expression for those generators in terms of
$V_{8I'}=(V_{8a},\ V_8{}^a,\ V_{87})$ and $V_{I'J'}=(V_{7a},\ V_7{}^a; V_{a}{}^b,\ V_a{}^4,\ V_4{}^a)$ is given by the relations \cite{U09}, \cite{U12} \footnote{In Ref.~\cite{GSWnew} the $so(8)$ generators adapted for the $S^7$ Hopf fibration  were given in conventional $D=6$ vector basis and
without specifying explicitly the K\" ahler 2-form on
$\mathbb{CP}^3$. However, natural realization of the $u(3)$ stability
algebra of $\mathbb{CP}^3$ is provided by the generators
$\widetilde V_a{}^b$ in $\mathbf 3\times\bar{\mathbf 3}$
representation of $SU(3)$. Associated representation of $D=6$ vectors in
$\mathbf 3\oplus\bar{\mathbf 3}$ basis suggests the
choice of the K\" ahler 2-form such that its contraction with the $su(4)$
generators constructed out of $D=6$ chiral gamma-matrices is given
by the diagonal $4\times 4$ matrix \cite{U10}.}
\beq
\begin{array}{c}
T_a=\frac12(V_{7a}-iV_{8a}),\quad T^a=-\frac12(V_7{}^{a}+iV_8{}^{a}),\quad\widetilde V_a{}^b=V_a{}^b-\frac12\delta_a^bV_c{}^c+\frac14\delta_a^bV_{87};\\[0.2cm]
H=V_{87}+2V_a{}^a;\quad\widetilde T_a=-\frac12(V_{7a}+iV_{8a}),\quad\widetilde T^a=\frac12(V_7{}^{a}-iV_8{}^{a}).
\end{array}
\eeq
Accordingly Cartan forms in this basis are expressed through $\Omega_{8I'}=(\Omega_{8a},\ \Omega_8{}^a,\ \Omega_{87})$ and $\Omega_{I'J'}=(\Omega_{7a},\ \Omega_7{}^a; \Omega_{a}{}^b,\ \Omega_a{}^4,\ \Omega_4{}^a)$ that enter (\ref{cf})
\beq
\begin{array}{c}
\Omega_a(d)=\Omega_{7a}-\frac{i}{2}\Omega_{8a},\quad\widetilde\Omega_a(d)=-\Omega_{7a}-\frac{i}{2}\Omega_{8a},\\[0.2cm]
\Omega^a(d)=-\Omega^{7a}-\frac{i}{2}\Omega^{8a},\quad\widetilde\Omega^a(d)=\Omega^{7a}-\frac{i}{2}\Omega^{8a},\\[0.2cm]
h(d)=\frac14(\Omega_{87}+2\Omega_a{}^a),\quad\widetilde\Omega_a{}^b(d)=\Omega_a{}^b-\frac12\delta_a^b\Omega_c{}^c+\frac14\delta_a^b\Omega_{87}.
\end{array}
\eeq
Finally $\mathcal C_{\mbox{\scriptsize 32susys}}$ is the linear combination of odd generators divided into those corresponding to unbroken $(Q^a_\mu,\ \bar Q_{\mu a};\ S^{\mu a},\ \bar S^\mu_a)$ and broken $(Q^4_\mu,\ \bar Q_{\mu 4};\ S^{\mu 4},\ \bar S^\mu_4)$ supersymmetries from the $AdS_4\times\mathbb{CP}^3$ superbackground perspective
\beq\label{fermibasis}
\begin{array}{rl}
\mathcal C_{\mbox{\scriptsize 32susys}}=&\underline\omega^\mu_a(d)Q^a_\mu+\underline{\bar\omega}{}^{\mu a}(d)\bar Q_{\mu a}+\omega^\mu_4(d)Q^4_\mu+\bar\omega^{\mu 4}(d)\bar Q_{\mu 4}\\[0.2cm]
+&\underline\chi{}_{\mu a}(d)S^{\mu a}+\underline{\bar\chi}{}^a_\mu(d)\bar S^\mu_a+\chi_{\mu 4}(d)S^{\mu 4}+\bar\chi^4_\mu(d)\bar S^\mu_4.
\end{array}
\eeq
Then the $D=11$ supervielbein bosonic components (\ref{11dvielgen}) are expressed through the Cartan forms (\ref{confbasis1}) and (\ref{hopfbasis})
\beq\label{11dbosvielcf}
\begin{array}{rl}
\hat E^{m'}(d)&=\left(\frac12(\underline{\omega}^m(d)+\underline c^m(d)),\ -\underline\Delta(d)\right),\\[0.2cm]
\hat E_{I'}(d)=&\left(i(\Omega_a(d)+\widetilde\Omega_a(d)),\ i(\Omega^a(d)+\widetilde\Omega^a(d)),\ h(d)+\widetilde\Omega_a{}^a(d)\right).
\end{array}
\eeq
Similarly supervielbein fermionic components coincide with the Cartan forms (\ref{fermibasis}).

Anticipating reduction to $D=10$ explicit expressions for the $AdS_4\times S^7$ supervielbein components can be conveniently written in terms of the $osp(4|6)$ Cartan forms, the differential of $S^1$ Hopf fiber coordinate $dy$ and 8 Grassmann coordinates $(\theta^\mu,\ \bar\theta^\mu,\ \eta_\mu,\ \bar\eta_\mu)$ for the supersymmetries broken by the $AdS_4\times\mathbb{CP}^3$ superbackground. The $OSp(4|8)/(SO(1,3)\times SO(7))$ representative $\hat{\mathscr G}$ can be taken in the form of a $OSp(4|6)/(SO(1,3)\times U(3))$ element 'dressed' by the $S^1$ Hopf fiber coordinate and 8 'broken' fermions, e.g. as in \cite{U09}, \cite{U12}
\beq\label{cosetrep}
\hat{\mathscr G}=\mathscr
Ge^{yH}e^{\theta^\mu
Q^4_{\mu}+\bar\theta^\mu\bar Q_{\mu 4}}e^{\eta_\mu S^{\mu
4}+\bar\eta_\mu\bar S^\mu_4}\in OSp(4|8)/(SO(1,3)\times SO(7))
\eeq
that ensures the supervielbein independence on $y$, however, contributions proportional to $dy$ do arise for fermionic components of the supervielbein and bosonic ones in directions tangent to the $AdS_4$ space-time because of the structure of underlying $osp(4|8)$ superalgebra.

Considerable simplification of the $AdS_4\times S^7$ supervielbein can be attained by partially fixing $\kappa-$symmetry gauge freedom in the broken supersymmetries sector. In \cite{U12} we have proposed the Kaluza-Klein $\kappa-$gauge characterized by vanishing of contributions proportional to $dy$ in the $D=11$ supervielbein bosonic components that are tangent to $D=10$ space-time. Another possibility of simplifying the $AdS_4\times S^7$ supergeometry is to set to zero Grassmann coordinates $\eta_\mu$, $\bar\eta_\mu$ associated with the broken part of $D=3$ $\mathcal N=8$ conformal supersymmetry. In such a partial $\kappa-$symmetry gauge the $AdS_4\times S^7$ supervielbein bosonic components acquire the form\footnote{In what follows concentrating on the particular choice of the $\kappa-$symmetry gauge we retain the same notation for the gauge-fixed expressions for $D=11$ and $D=10$ supervielbein components.}
\beq\label{11dviel}
\begin{array}{rl}
\hat E^m(d)=&\mathfrak e^m(d)+\frac14c^m(d)\theta^2\bar\theta^2+dyG^{m}_y,\quad\hat
E^{11}(d)=dy+\widetilde\Omega_{a}{}^a(d)+c^m(d)(\theta\sigma_m\bar\theta);\\[0.2cm]
E^3(d)=&-\Delta(d),\quad E_a(d)=i(\Omega_a(d)+2\chi_{\mu a}(d)\theta^\mu),\quad E^a(d)=i(\Omega^a(d)-2\bar\chi_\mu^a(d)\bar\theta^\mu),
\end{array}
\eeq
where
\beq
\mathfrak e^m(d)=G^{0'm}(d)-\frac{i}{2}\left(d\theta\sigma^m\bar\theta+d\bar\theta\sigma^m\theta+\varepsilon^{mkl}G_{kl}(d)(\theta\bar\theta)\right)
\eeq
and
\beq
G^{m}_y=2(\theta\sigma^m\bar\theta).
\eeq
We adopt the following notation for contractions of 2-component spinors $(\theta\sigma^m\bar\theta)=\theta^\mu\sigma^m_{\mu\nu}\bar\theta^\nu$, $(\theta\bar\theta)=\theta^\mu\bar\theta_\mu$ etc. The presence of contributions proportional to $dy$ necessitates the $SO(1,3)$ Lorentz rotation to be performed in directions tangent to the Minkowski boundary of $AdS_4$ and $S^1$
\beq
\begin{array}{rl}
(L\hat E)^{m}(d)=& L^{m}{}_{n}\hat E^{n}+L^{m}{}_{11}\hat E^{11}\\[0.2cm]
(L\hat E)^{11}(d)=& L^{11}{}_{m}\hat E^{m}+L^{11}{}_{11}\hat E^{11}=\Phi_L(dy+A_L),
\end{array}
\eeq
where $\Phi_L=e^{2\phi/3}$ is related to the $D=10$ dilaton $\phi$ and $A_L$ is identified with the RR 1-form potential. Lorentz rotation matrix
\beq\label{Lrot}
||L||=\left(
\begin{array}{rl}
L^{m}{}_{n} & L^{m}{}_{11}\\[0.2cm]
L^{11}{}_{m} & L^{11}{}_{11}
\end{array}
\right)\in SO(1,3)
\eeq
is defined by the requirement of removing the summand proportional to $dy$ in $(L\hat E)^{m}$, i.e.
$L^{m}{}_{n}G^{n}_y+L^{m}{}_{11}=0$. Explicit form of its entries is as follows
\beq\label{Lrot2}
\begin{array}{rl}
L^{m}{}_{n}=&(1-\theta^2\bar\theta^2)\delta^m_n,\quad
L^{ m}{}_{11}=-2(\theta\sigma^m\bar\theta),\\[0.4cm]
L^{11}{}_{m}=&2(\theta\sigma_m\bar\theta),\quad L^{11}{}_{11}=1-3\theta^2\bar\theta^2.
\end{array}
\eeq
Then $(L\hat E)^{m}$ is identified with the $D=10$ supervielbein bosonic components $E^{m}$ in directions tangent to the Minkowski boundary of $AdS_4$ space-time
\beq\label{3compviel}
E^m(d)=\mathfrak e^m(d)-2\widetilde\Omega_{a}{}^a(d)(\theta\sigma^m\bar\theta)-\frac74G^{0'm}(d)\theta^2\bar\theta^2-\frac34G^{3m}(d)\theta^2\bar\theta^2.
\eeq
The scalar superfield is given by the following expression
\beq\label{defphi}
\Phi_L=1+3\theta^2\bar\theta^2.
\eeq
Other bosonic components of the $D=11$ supervielbein $E^3(d)$, $E_a(d)$ and $E^a(d)$ are not affected by the Lorentz rotation and can be directly identified with the $D=10$ supervielbein bosonic components in the corresponding $\kappa-$symmetry gauge\footnote{Fermionic supervielbein components are transformed under the spinor version of the above Lorentz rotation (see \cite{GSWnew}, \cite{GrSW}, \cite{U09}). We do not discuss it here since the superparticle action is constructed out the $D=10$ supervielbein bosonic components only.}.

\subsection{$AdS_4\times\mathbb{CP}^3$ superparticle action and equations of motion in the partial $\kappa-$symmetry gauge}

Superparticle action on the $AdS_4\times\mathbb{CP}^3$ superbackground
\beq\label{action_ini}
\mathscr S=\int\frac{d\tau}{e}\Phi_L\left(E_{\tau m}E^{m}_\tau+E^3_\tau E^3_\tau-E_{\tau a}E_\tau\vp{E}^a\right)
\eeq
in considered partial $\kappa-$symmetry gauge by substituting expressions for the gauge-fixed supervielbein bosonic components (\ref{3compviel}) and (\ref{11dviel}) is brought to the form
\beq\label{action}
\begin{array}{rl}
\mathscr S_{\mbox{\scriptsize g.f.}}=&{\displaystyle\int\frac{d\tau}{e}}\left[\mathfrak e_{\tau m}\mathfrak e^m_\tau-4\widetilde\Omega_{\tau a}{}^a\mathfrak e_{\tau m}(\theta\sigma^m\bar\theta)+6\widetilde\Omega_{\tau a}{}^a\widetilde\Omega_{\tau b}{}^b\theta^2\bar\theta^2-\frac12G_\tau\vp{G}^{0'}\vp{G}_mG_\tau\vp{G}^{0'm}\theta^2\bar\theta^2\right.\\[0.3cm]
-&\left.\frac32G_\tau\vp{G}^{0'}\vp{G}_mG_\tau\vp{G}^{3m}\theta^2\bar\theta^2+\Phi_L(\Delta_\tau\Delta_\tau-E_{\tau a}E_\tau\vp{E}^a)\right].
\end{array}
\eeq
To derive equations of motion following from the action (\ref{action}) analogously to the case of superparticle on the $OSp(4|6)/(SO(1,3)\times U(3))$ supercoset manifold it is helpful to consider as variation parameters the $osp(4|6)$ Cartan forms with definite $\mathbb Z_4-$grading (\ref{gradedcf}). Also we set the Lagrange multiplier $e$ to unity since the mass-shell constraint is not involved in the Lax representation for the equations of motion. In such a way it is possible to get the set of bosonic equations of motion
\beq\label{boseom}
\begin{array}{rl}
-\frac12\frac{\delta\mathscr S_{\mbox{\scriptsize g.f.}}}{\delta G^{0'}{}_m(\delta)}=&\frac{d}{d\tau}\mathfrak E_{\tau}{}^m+2G_{\tau}\vp{G}^m\vp{G}_n\mathfrak E_{\tau}\vp{\mathfrak E}^{n}+2\Phi_LG_\tau\vp{G}^{3m}\Delta_\tau+iE_{\tau a}(\bar\omega^a_\tau\sigma^m\bar\theta)-iE_\tau\vp{E}^a(\omega_{\tau a}\sigma^m\theta)\\[0.2cm]
+&2i\varepsilon^{mkl}\mathfrak e_{\tau k}G_\tau\vp{G}^{0'}\vp{G}_l(\theta\bar\theta)-\frac32\Delta_\tau G_\tau\vp{G}^{0'm}\theta^2\bar\theta^2=0,\\[0.2cm]
-\frac12\frac{\delta\mathscr S_{\mbox{\scriptsize g.f.}}}{\delta\Delta(\delta)}=&\frac{d}{d\tau}\left(\Phi_L\Delta_\tau\right)-2G_{\tau3m}\mathfrak E_{\tau}{}^{m}+iE_{\tau a}\bar\chi^{\vp{\mu a}}_\tau\vp{\bar\chi}^{\mu a}\bar\theta_\mu-iE_\tau\vp{E}^a\chi^{\vp{\mu}}_\tau\vp{\chi}^\mu_a\theta_\mu\\[0.2cm] +&\frac32G_\tau\vp{G}^{0'}\vp{G}_mG_\tau\vp{G}^{0'm}\theta^2\bar\theta^2=0,\\[0.2cm]
-i\frac{\delta\mathscr S_{\mbox{\scriptsize g.f.}}}{\delta\Omega_a(\delta)}=&\frac{d}{d\tau}\left(\Phi_L E_\tau\vp{E}^a\right)+i\Phi_L E_\tau\vp{E}^b(\widetilde\Omega_{\tau b}\vp{\widetilde\Omega}^a+\delta_b^a\widetilde\Omega_{\tau c}\vp{\widetilde\Omega}^c)-4\Omega_{\tau}\vp{\Omega}^a\mathfrak E_{\tau}{}^{m}(\theta\sigma_m\bar\theta)\\[0.2cm]
-&2i\varepsilon^{abc}E_{\tau b}\chi^{\vp{\mu}}_\tau\vp{\chi}^\mu_c\bar\theta_\mu=0,\\[0.2cm]
-i\frac{\delta\mathscr S_{\mbox{\scriptsize g.f.}}}{\delta\Omega^a(\delta)}=&\frac{d}{d\tau}\left(\Phi_L E_{\tau a}\right)-i(\widetilde\Omega_{\tau a}\vp{\widetilde\Omega}^b+\delta^b_a\widetilde\Omega_{\tau c}\vp{\widetilde\Omega}^c)\Phi_L E_{\tau b}+4\Omega_{\tau a}\mathfrak E_{\tau}{}^{m}(\theta\sigma_m\bar\theta)\\[0.2cm]
-&2i\varepsilon_{abc}E_{\tau}\vp{E}^{b}\bar\chi^{\vp{\mu}}_\tau\vp{\bar\chi}^{\mu c}\theta_\mu=0,\\[0.2cm]
\end{array}
\eeq
where
\beq
\mathfrak E_{\tau}{}^m=\Phi_LE_{\tau}{}^m-\frac74G_\tau{}^{0'm}\theta^2\bar\theta^2
%\mathfrak e_{\tau}{}^{m}-2\widetilde\Omega_{\tau a}{}^a(\theta\sigma^m\bar\theta)-\frac12G_\tau{}^{0'm}\theta^2\bar\theta^2
%-\frac34G_\tau{}^{3m}\theta^2\bar\theta^2
\eeq
and $D=10$ supervielbein components tangent to the $\mathbb{CP}^3$ manifold $E_a(d)$, $E^a(d)$ are given in (\ref{11dviel}).
Analogously fermionic equations arise from the variation w.r.t. $\mathbb Z_4-$graded fermionic $osp(4|6)$ Cartan forms
\beq
\begin{array}{rl}
-\frac14\frac{\delta\mathscr S_{\mbox{\scriptsize g.f.}}}{\delta\bar\omega_{(1)}\vp{\bar\omega}_\mu^a(\delta)}=&i\mathfrak E_{\tau}{}^{m}\tilde\sigma^{\mu\nu}_m\omega_{(1)\tau\nu a}+\Phi_L\left[\Delta_\tau\omega^{\vp{\nu}}_{(1)}\vp{\omega}^{\vp{\mu}}_{\tau}\vp{\omega}^{\mu}_a-\varepsilon_{abc}E_\tau\vp{E}^b(\bar\omega_{(1)\tau}\vp{\bar\omega}^{\mu c}+i\theta^\mu\bar\chi\vp{\chi}^{\vp{c}}_\tau\vp{\chi}^c_\nu\bar\theta^\nu)\right]\\[0.2cm]
+&\frac{1}{2}\frac{d}{d\tau}\left(E_{\tau a}\bar\theta^\mu\right)-\frac{i}{2}\left(\widetilde\Omega_{\tau a}\vp{\widetilde\Omega}^b-\delta_a^b\widetilde\Omega_{\tau c}\vp{\widetilde\Omega}^c\right)E_{\tau b}\bar\theta^\mu+\frac{1}{4}E_{\tau a}\left[G_\tau\vp{G}^{mn}\bar\theta^\nu\sigma_{mn\nu}{}^\mu\right.\\[0.2cm]
+&\left.2\Delta_\tau\bar\theta^\mu+2i(G_\tau\vp{G}^{0'm}+G_\tau\vp{G}^{3m})\tilde\sigma^{\mu\nu}_m\bar\theta_\nu\right]-i\mathfrak e_\tau{}^m\tilde\sigma^{\mu\nu}_m\omega^{\vp{\nu}}_{(3)\tau\nu a}(\theta\bar\theta)\\[0.2cm]
-&2i\omega_{(3)}\vp{\omega}^{\vp{\mu}}_{\tau}\vp{\omega}^\mu_a\mathfrak E_{\tau}{}^{m}(\theta\sigma_m\bar\theta)+\frac{3i}{4}G_\tau\vp{G}^{0'm}\tilde\sigma^{\mu\nu}_m\omega^{\vp{\nu}}_{(3)\tau\nu a}\theta^2\bar\theta^2=0;
\end{array}
\eeq
\beq\label{fermeom}
\begin{array}{rl}
-\frac14\frac{\delta\mathscr S_{\mbox{\scriptsize g.f.}}}{\delta\bar\omega_{(3)}\vp{\bar\omega}_\mu^a(\delta)}=&i\mathfrak E_{\tau}{}^{m}\tilde\sigma^{\mu\nu}_m\omega_{(3)\tau\nu a}-\Phi_L\left[\Delta_\tau\omega^{\vp{\nu}}_{(3)}\vp{\omega}^{\vp{\mu}}_{\tau}\vp{\omega}^{\mu}_a-\varepsilon_{abc}E_\tau\vp{E}^b(\bar\omega_{(3)\tau}\vp{\bar\omega}^{\mu c}+i\theta^\mu\bar\chi\vp{\chi}^{\vp{c}}_\tau\vp{\chi}^c_\nu\bar\theta^\nu)\right]\\[0.2cm]
-&\frac{1}{2}\frac{d}{d\tau}\left(E_{\tau a}\bar\theta^\mu\right)+\frac{i}{2}\left(\widetilde\Omega_{\tau a}\vp{\widetilde\Omega}^b-\delta_a^b\widetilde\Omega_{\tau c}\vp{\widetilde\Omega}^c\right)E_{\tau b}\bar\theta^\mu-\frac{1}{4}E_{\tau a}\left[G_\tau\vp{G}^{mn}\bar\theta^\nu\sigma_{mn\nu}{}^\mu\right.\\[0.2cm]
+&\left.2\Delta_\tau\bar\theta^\mu-2i(G_\tau\vp{G}^{0'm}+G_\tau\vp{G}^{3m})\tilde\sigma^{\mu\nu}_m\bar\theta_\nu\right]+i\mathfrak e_\tau{}^m\tilde\sigma^{\mu\nu}_m\omega^{\vp{\nu}}_{(1)\tau\nu a}(\theta\bar\theta)\\[0.2cm]
+&2i\omega_{(1)}\vp{\omega}^{\vp{\mu}}_{\tau}\vp{\omega}^\mu_a\mathfrak E_{\tau}{}^{m}(\theta\sigma_m\bar\theta)+\frac{3i}{4}G_\tau\vp{G}^{0'm}\tilde\sigma^{\mu\nu}_m\omega^{\vp{\nu}}_{(1)\tau\nu a}\theta^2\bar\theta^2=0.
\end{array}
\eeq
Observe that Eqs.~(\ref{boseom})-(\ref{fermeom}) reduce to the equations of motion (\ref{coseteom}) for the superparticle on the $OSp(4|6)/(SO(1,3)\times U(3))$ supermanifold when Grassmann coordinates $\theta^\mu$, $\bar\theta^\mu$ are set to zero. There are also 8 fermionic equations associated with the broken supersymmetries
\beq
\begin{array}{rl}
\frac{\delta\mathscr S_{\mbox{\scriptsize g.f.}}}{\vp{\hat{\bar\Psi}}\delta\bar\theta_\mu}=&-2iE_{\tau a}\bar\chi^{\vp{\mu a}}_\tau\vp{\chi}^{\mu a}-i\frac{d}{d\tau}\left(\mathfrak E_\tau{}^m\tilde\sigma^{\mu\nu}_m\theta_\nu\right)-i\mathfrak e_\tau{}^m\left(\tilde\sigma^{\mu\nu}_m\dot\theta_\nu+\theta^\mu\varepsilon_{mkl}G_\tau{}^{kl}\right)\\[0.2cm]
-&4\widetilde\Omega_{\tau a}{}^a\left[G_\tau\vp{G}^{0'm}-2\widetilde\Omega_{\tau b}{}^b(\theta\sigma^m\bar\theta)\right]\tilde\sigma^{\mu\nu}_m\theta_\nu+6\bar\theta^\mu(\Delta_\tau\Delta_\tau+\Omega_{\tau a}\Omega_\tau{}^a)\theta^2\\[0.2cm]
-&\bar\theta^\mu G_\tau\vp{G}^{0'}\vp{G}_m(G_\tau\vp{G}^{0'm}+3G_\tau\vp{G}^{3m})\theta^2-3i\widetilde\Omega_{\tau a}{}^a\left(\dot{\bar\theta}{}^\mu\theta^2-2\bar\theta^\mu(\dot\theta\theta)\right)=0;
\end{array}
\eeq
\beq\label{etaeom}
\begin{array}{rl}
-&2i\Phi_LE_{\tau a}\left[\bar\omega_\tau\vp{\bar\omega}^{\mu a}+i\bar\chi_\tau\vp{\bar\chi}^{\mu a}(\theta\bar\theta)\right]-2i\mathfrak e_\tau{}^mc_\tau{}^n(\theta\sigma_n\tilde\sigma_m)^\mu+2i\Delta_\tau\left(\dot\theta^\mu-2i\widetilde\Omega_{\tau a}{}^a\theta^\mu\right.\\[0.2cm]
+&\left.\frac12G_\tau\vp{G}^{mn}\theta^\nu\sigma_{mn\nu}{}^\mu\right)-2i\theta^\mu\left(\mathfrak e_{\tau m}\mathfrak e_\tau{}^m-E_{\tau a}E_\tau\vp{E}^a\right)+(\Delta_\tau+2i\widetilde\Omega_{\tau a}{}^a)c_\tau{}^m\tilde\sigma^{\mu\nu}_m\bar\theta_\nu\theta^2\\[0.2cm]
+&\!4\mathfrak E_{\tau}{}^{m}(\theta\sigma_m\bar\theta)(\dot\theta^\mu\!+\!\frac12G_\tau\vp{G}^{kl}\theta^\nu\sigma_{kl\nu}{}^\mu)\!+\!6\Delta_\tau G_\tau\vp{G}^{0'm}\tilde\sigma^{\mu\nu}_m\bar\theta_\nu\theta^2\!+\!\frac{21i}{2}\Delta_\tau\dot\theta^\mu\theta^2\bar\theta^2=0
\end{array}
\eeq and c.c. When $\theta^\mu=\bar\theta^\mu=0$ these equations
coincide with (\ref{orthogeom}).

\subsection{Lax representation for $AdS_4\times\mathbb{CP}^3$ superparticle equations of motion in the partial $\kappa-$symmetry gauge}

Similarly to equations of motion (\ref{coseteom}) for the
superparticle on the $OSp(4|6)/(SO(1,3)\times U(3))$ supercoset
space the system (\ref{boseom})-(\ref{etaeom}) also admits Lax
representation \beq\label{lax1} \frac{dL}{d\tau}+[M,L]=0 \eeq with
$M$ being the same as in (\ref{cosetlax}) and $L$ now taking value
in the whole $osp(4|6)$ superalgebra rather than its $g_{(2)}$
eigenspace
\beq\label{lax2}
L=L_{\mbox{\scriptsize
so(2,3)}}+L_{\mbox{\scriptsize
su(4)}}+L_{\mbox{\scriptsize24susys}}\in osp(4|6).
\eeq
The first
term describes anti-de Sitter part of $L$
\beq\label{lax2so}
L_{\mbox{\scriptsize
so(2,3)}}=2\!\left(\!\Phi_LE_{\tau}{}^{m}\!-\!\frac74\theta^2\bar\theta^2G_\tau{}^{0'm}\!\right)\!M_{0'm}\!+\Phi_L\Delta_\tau
D+\!\frac32\theta^2\bar\theta^2G_\tau{}^{0'm}M_{3m}\!+i(\theta\bar\theta)\varepsilon^{kmn}\mathfrak
e_{\tau k}M_{mn}.
\eeq
The second summand in
(\ref{lax2}) belongs to the $su(4)$ isometry algebra of
$\mathbb{CP}^3$ manifold
\beq\label{lax2su}
L_{\mbox{\scriptsize
su(4)}}=-i\Phi_L(E_{\tau
a}T^a+E_\tau\vp{E}^aT_a)-4(\theta\sigma_m\bar\theta)E_{\tau}{}^{m}\widetilde
V_a{}^a.
\eeq
The contribution proportional to the
$osp(4|6)$ fermionic generators
\beq
L_{\mbox{\scriptsize24susys}}=\varepsilon_{(1)}{}^\mu_{a}Q^{\vp{a}}_{(1)}\vp{Q}^a_\mu+\bar\varepsilon_{(1)}{}^{\mu
a}\bar Q_{(1)\mu
a}+\varepsilon_{(3)}{}^\mu_{a}Q^{\vp{a}}_{(3)}\vp{Q}^a_\mu+\bar\varepsilon_{(3)}{}^{\mu
a}\bar Q_{(3)\mu a},
\eeq
where
\beq
\varepsilon_{(1)}{}^\mu_{a}=\varepsilon_{(3)}{}^\mu_{a}=-\frac{1}{2}E_{\tau
a}\bar\theta^\mu,\quad\bar\varepsilon_{(1)}{}^{\mu
a}=\bar\varepsilon_{(3)}{}^{\mu
a}=\frac{1}{2}E^{\vp{a}}_{\tau}\vp{E}^a_{\vp{tau}}\theta^\mu, \eeq
by using (\ref{gradedfermidef}) can be written simply as \beq
L_{\mbox{\scriptsize24susys}}=-E_{\tau a}\bar\theta^\mu
Q^a_\mu+E^{\vp{a}}_{\tau}\vp{E}^a_{\vp{tau}}\theta^\mu\bar Q_{\mu
a}.
\eeq

Lax connection (\ref{lax2}) can be presented as an $osp(4|6)-$valued differential operator acting on the superparticle action (\ref{action})
\beq\label{lax2compl}
\begin{array}{rl}
L=&\left(M_{0'm}\frac{\partial}{\partial G_{\tau}\vp{G}^{0'}\vp{G}_m}+\frac12D\frac{\partial}{\partial\Delta_{\tau}}-M_{mn}\frac{\partial}{\partial G_{\tau mn}}-M_{3m}\frac{\partial}{\partial G_{\tau 3m}}+T^a\frac{\partial}{\partial\Omega_{\tau}\vp{\Omega}^a}+T_a\frac{\partial}{\partial\Omega_{\tau a}}+\widetilde V_a{}^a\frac{\partial}{\partial\widetilde\Omega_{\tau b}{}^b}\right.\\[0.2cm]
-&\left.\frac14Q^{\vp{a}}_{(1)}\vp{Q}^a_\mu\frac{\partial}{\partial\bar\omega^{\vp{a}}_{(3)}\vp{\bar\omega}^{\vp{a}}_{\tau}\vp{\bar\omega}^a_\mu}+\frac14\bar Q_{(1)\mu a}\frac{\partial}{\partial\omega_{(3)\tau\mu a}}+\frac14Q^{\vp{a}}_{(3)}\vp{Q}^a_\mu\frac{\partial}{\partial\bar\omega^{\vp{a}}_{(1)}\vp{\bar\omega}^{\vp{a}}_{\tau}\vp{\bar\omega}^a_\mu}-\frac14\bar Q_{(3)\mu a}\frac{\partial}{\partial\omega_{(1)\tau\mu a}}\right)\mathscr S_{\mbox{\scriptsize g.f.}}
\end{array}
\eeq
generalizing corresponding expression (\ref{diffoper}) for the $OSp(4|6)/(SO(1,3)\times U(3))$ superparticle. We suggest that complete equations of motion for the $AdS_4\times\mathbb{CP}^3$ superparticle that take into account all 8 fermionic coordinates related to the broken space-time supersymmetries are encoded in the same Lax representation (\ref{lax1}) with $L$ given by (\ref{lax2compl}), where instead of $\mathscr S_{\mbox{\scriptsize g.f.}}$ the complete action (\ref{action_ini}) should be substituted.

Let us mention that one can use the freedom in the definition of the Lagrange multiplier $e(\tau)$ in the superparticle action (\ref{action_ini}) to 'absorb' $\Phi_L$. In the considered partial $\kappa-$symmetry gauge this amounts to the following substitution
\beq
-\frac12G_\tau\vp{G}^{0'}\vp{G}_mG_\tau\vp{G}^{0'm}\theta^2\bar\theta^2\rightarrow-\frac72G_\tau\vp{G}^{0'}\vp{G}_mG_\tau\vp{G}^{0'm}\theta^2\bar\theta^2
\eeq 
in the action (\ref{action}). Corresponding equations of motion and the Lax pair are obtained from (\ref{boseom})-(\ref{fermeom}) and (\ref{lax2})-(\ref{lax2su}) by setting $\Phi_L=1$.

It is interesting to examine how integrable structure of the $AdS_4\times\mathbb{CP}^3$ superparticle is related to that of the superstring. The Lax connection of the $AdS_4\times\mathbb{CP}^3$ superstring $\mathcal L_i$, $i=(\tau,\,\sigma)$, is given by the sum
\beq\label{lax-string}
\mathcal L_i=\widehat{\mathcal L}_i+*\widetilde{\mathcal L}_i,
\eeq
where the last term contains $2d$ Hodge dual of $\widetilde{\mathcal L}_i$: $*\widetilde{\mathcal L}_i=\varepsilon_{ij}\gamma^{jk}\widetilde{\mathcal L}_k=\gamma_{ij}\varepsilon^{jk}\widetilde{\mathcal L}_k$. Both summands depend on the parameters $\ell_1$, $\ell_2$, $\ell_3$ and $\ell_4$ that are the same as those entering the Lax connection of $OSp(4|6)/(SO(1,3)\times U(3))$ sigma-model \cite{AF}, \cite{Stefanski}. They satisfy the constraints
\beq
\ell^2_1-\ell^2_2=\ell_3\ell_4=1,\quad(\ell_1-\ell_2)\ell_4=\ell_3,\quad(\ell_1+\ell_2)\ell_3=\ell_4
\eeq
that can be solved to recover dependence on a single spectral parameter, e.g. as follows
\beq
\ell_1=\frac12\left(\frac{1}{z^2}+z^2\right),\quad \ell_2=\frac12\left(\frac{1}{z^2}-z^2\right),\quad\ell_3=z,\quad\ell_4=\frac{1}{z}.
\eeq
The Lax connection (\ref{lax-string}) satisfies $2d$ zero curvature condition
\beq\label{zerocur}
\partial_\sigma\mathcal L_\tau-\partial_\tau\mathcal L_\sigma-[\mathcal L_\tau,\mathcal L_\sigma]=0.
\eeq The superparticle limit amounts to dropping the dependence of
the superspace coordinate fields on the world-sheet space-like
coordinate $\sigma$ and taking $z$ to unity. Since the Lax
connection carries $2d$ vector index only due to the world-sheet
derivatives of the superspace coordinate fields it follows that
$\widehat{\mathcal L}_\sigma=*\widetilde{\mathcal L}_\tau=0$. Using
that $\widetilde{\mathcal L}_i$ has overall factor of $\ell_2$
\cite{SW10}, \cite{CSW} one can perform the rescaling and define
\beq 
L=\lim\limits_{z\to1}\frac{1}{\ell_2}*\widetilde{\mathcal
L}_\sigma,\quad M=\lim\limits_{z\to1}\widehat{\mathcal L}_\tau 
\eeq
so that (\ref{zerocur}) transforms into the Lax representation
(\ref{lax1}) for the $AdS_4\times\mathbb{CP}^3$ superparticle
equations of motion. Reasoning in the opposite direction we
conclude that, once the Lax pair for the superparticle is known,
this allows to recover dependence of the superstring Lax
connection on the superspace coordinates (but not the dependence
on the spectral parameter) up to terms linear in $\ell_2$ for
$\widehat{\mathcal L}_i$ and quadratic in $\ell_2$ for
$\widetilde{\mathcal L}_i$. For $\widehat{\mathcal L}_i$ this just
yields the corresponding part of $OSp(4|6)/(SO(1,3)\times U(3))$
sigma-model Lax connection since the terms containing Grassmann
coordinates related to the broken supersymmetries are proportional
to $\ell_2$ \cite{SW10}, \cite{CSW}, while for
$\widetilde{\mathcal L}_i$ the superparticle limit preserves
non-trivial information on its dependence on the 'broken'
fermions.

\section{Conclusion}

For superstring models consistent zero-mode limit can be defined that is described by massless superparticles. Superparticle mechanics is in general easier to analyze than that of the corresponding superstring model since there survives only finite number of degrees of freedom. That is why in this paper we addressed Lagrangian mechanics of the massless superparticle on the $OSp(4|6)/(SO(1,3)\times U(3))$ supercoset manifold and $AdS_4\times\mathbb{CP}^3$ superbackground. Because all $OSp(4|6)/(SO(1,3)\times U(3))$ supercoset coordinates can be associated with the (super)symmetries of the manifold, superparticle equations of motion, although non-linear, are expressed in terms of the  $osp(4|6)$ Cartan forms and admit the Lax representation implying their integrability.

The case of $AdS_4\times\mathbb{CP}^3$ superparticle is more
involved since the $AdS_4\times\mathbb{CP}^3$ superspace is
additionally parametrized by 8 anticommuting coordinates for the
broken supersymmetries and is not isomorphic to a supercoset
manifold. It is not apriori obvious that integrable structure of
the superparticle/superstring can be extended from the
$OSp(4|6)/(SO(1,3)\times U(3))$ supermanifold into the
$AdS_4\times\mathbb{CP}^3$ superspace. First evidence that this is
indeed so was given in \cite{SW10}, \cite{CSW}, where the Lax
connection of the $OSp(4|6)/(SO(1,3)\times U(3))$ sigma-model had
been extended by linear and quadratic contributions of all 8
fermionic coordinates associated with the broken supersymmetries.
In Ref.~\cite{U12} we proposed to use the $\kappa-$symmetry gauge
freedom to keep in the broken supersymmetries sector less degrees
of freedom that facilitates study of the integrable structure
beyond the $OSp(4|6)/(SO(1,3)\times U(3))$ supermanifold. 

In the present pape we
considered the case when 4 fermionic coordinates related to broken
part of conformal supersymmetry are gauged away so that there
remain 4 coordinates corresponding to broken Poincare
supersymmetry. Equations of motion for the
$AdS_4\times\mathbb{CP}^3$ superparticle in such a partial
$\kappa-$gauge have been derived and shown to be integrable. The
Lax pair includes contributions up to the 4th order in the
'broken' fermions. It can be expressed in terms of the
differential operator that takes value in the $osp(4|6)$
superalgebra and acts of the superparticle action. We suggest that
this form of the Lax pair is generic one for the
$AdS_4\times\mathbb{CP}^3$ superparticle, i.e. does not depend on
a particular $\kappa-$symmetry gauge choice. Proving it would
establish the integrability of the $AdS_4\times\mathbb{CP}^3$
superparticle and provide further evidence in favor of the
integrability of complete equations of motion for the $AdS_4\times\mathbb{CP}^3$
superstring that is important for identifying its
spectrum. Finally we explored the relation between the integrable
structures of the superparticle and superstring and found that from
the Lax pair of the $AdS_4\times\mathbb{CP}^3$ superparticle one
can gain information on the Hodge dualized part of the superstring
Lax connection.

\section{Acknowledgements}

The author is grateful to A.A.~Zheltukhin for stimulating discussions.


\begin{thebibliography}{99}
\bibitem{ABJM}
O.~Aharony, O.~Bergman, D.L.~Jafferis and J.~Maldacena, "$\mathcal
N=6$ superconformal Chern-Simons-matter theories, M2-branes and
their gravity duals", JHEP \textbf{0810} (2008) 091,
\href{http://arxiv.org/abs/0806.1218}{\texttt{arXiv:0806.1218
[hep-th]}}.

\bibitem{Maldacena}
J.M.~Maldacena, "The large N limit of superconformal field theories and supergravity", Adv.\ Theor.\ Math.\ Phys.\ \textbf{2} (1998) 231, \href{http://arxiv.org/abs/hep-th/9711200}{\texttt{arXiv:hep-th/9711200}}.\\
S.S.~Gubser, I.R.~Klebanov and A.M.~Polyakov, "Gauge theory correlators from non-critical string theory", Phys.\ Lett. \textbf{B428} (1998) 105, \href{http://arxiv.org/abs/hep-th/9802109}{\texttt{arXiv:hep-th/9802109}}.\\
E.~Witten, "Anti-de Sitter space and holography", Adv.\ Theor.\ Math.\ Phys.\ \textbf{2} (1998) 253, \href{http://arxiv.org/abs/hep-th/9802150}{\texttt{arXiv:hep-th/9802150}}.

\bibitem{Watamura}
S.~Watamura, "Spontaneous compactification and $Cp(N)$: $SU(3)\times SU(2)\times U(1)$, $\sin^2{\theta_W}$, $g(3)/g(2)$ and $SU(3)$ triplet chiral fermions in 4 dimensions", Phys. Lett. \textbf{B136} (1984) 245.

\bibitem{NPope}
B.E.W.~Nilsson and C.~Pope, "Hopf fibration of eleven dimensional supergravity", Class. Quantum Grav. \textbf{1} (1984) 499.

\bibitem{STV}
D.P.~Sorokin, V.I.~Tkach and D.V.~Volkov, "Kaluza-Klein theories
and spontaneous compactification mechanisms of extra space
dimensions", \emph{In *Moscow 1984, Proceedings, Quantum Gravity*,
376-392}.\\
D.P.~Sorokin, V.I.~Tkach and D.V.~Volkov, "On the
relationship between compactified vacua of $D=11$ and $D=10$
supergravities", Phys. Lett. \textbf{B161} (1985) 301.

\bibitem{MT98}
R.R.~Metsaev and A.A.~Tseytlin, "Type IIB superstring action in
$AdS_5\times S^5$ background", Nucl.\ Phys.\ \textbf{B533} (1998)
109, \href{http://arxiv.org/abs/hep-th/9805028}{\texttt{arXiv:hep-th/9805028}}.

\bibitem{AF}
G.~Arutyunov and S.~Frolov, "Superstrings on $AdS_4\times
CP^3$ as a Coset Sigma-model", JHEP \textbf{0809} (2008) 129, \href{http://arxiv.org/abs/0806.4940}{\texttt{arXiv:0806.4940 [hep-th]}}.

\bibitem{Stefanski}
B.J.~Stefanski, "Green-Schwarz action for Type IIA
strings on $AdS_4\times CP^3$", Nucl.\ Phys.\ \textbf{B808} (2009)
80, \href{http://arxiv.org/abs/0806.4948}{\texttt{arXiv:0806.4948 [hep-th]}}.

\bibitem{PS}
P.~Fre and P.A.~Grassi, "Pure Spinor Formalism for
$OSp(N|4)$ backgrounds", \href{http://arxiv.org/abs/0807.0044}{\texttt{arXiv:0807.0044 [hep-th]}}.\\
G.~Bonelli, P.A.~Grassi and H.~Safaai, "Exploring pure spinor string theory on $AdS_4\times\mathbb{CP}^3$", JHEP \textbf{0810} (2008) 085, \href{http://arxiv.org/abs/0808.1051}{\texttt{arXiv:0808.1051 [hep-th]}}.\\
R.~D'Auria, P.~Fre, P.A.~Grassi and M.~Trigiante,
"Superstrings on $AdS_4\times CP^3$ from Supergravity", Phys. Rev. \textbf{D79} (2009) 086001,
\href{http://arxiv.org/abs/0808.1282}{\texttt{arXiv:0808.1282 [hep-th]}}.

\bibitem{DHIS}
M.J.~Duff, P.S.~Howe, T.~Inami and K.S.~Stelle, "Superstrings in
$D=10$ from supermembranes in $D=11$", Phys. Lett. \textbf{B191}
(1987) 70.

\bibitem{HoweSezgin}
P.S.~Howe and E.~Sezgin, "The supermembrane revisited", Class. Quantum Grav. \textbf{22} (2005) 2167, \href{http://arxiv.org/abs/hep-th/0412245}{\texttt{arXiv:hep-th/0412245}}.

\bibitem{de Wit}
B.~de Wit, K.~Peeters, J.~Plefka and A.~Sevrin, "The M-theory two-brane in $AdS_4\times S^7$ and $AdS_7\times S^4$", Phys. Lett. \textbf{B443} (1998) 153, \href{http://arxiv.org/abs/hep-th/9808052}{\texttt{arXiv:hep-th/9808052}}.

\bibitem{GSWnew}
J.~Gomis, D.~Sorokin and L.~Wulff, "The complete $AdS_4\times
CP^3$ superspace for type IIA superstring and $D-$branes", JHEP
\textbf{0903} (2009) 015, \href{http://arxiv.org/abs/0811.1566}{\texttt{arXiv:0811.1566 [hep-th]}}.

\bibitem{GrSW}
P.A.~Grassi, D.~Sorokin and L.~Wulff, "Simplifying superstring and
$D-$brane actions in $AdS_4\times\mathbb{CP}^3$ superbackground", JHEP
\textbf{0908} (2009) 060,
\href{http://arxiv.org/abs/0903.5407}{\texttt{arXiv:0903.5407
[hep-th]}}.

\bibitem{U09}
D.V.~Uvarov, "$AdS_4\times\mathbb{CP}^3$ superstring in the light-cone gauge", Nucl. Phys. \textbf{B826} (2010) 294, \href{http://arxiv.org/abs/0906.4699}{\texttt{arXiv:0906.4699 [hep-th]}}.\\
D.V.~Uvarov, "Light-cone gauge Hamiltonian for $AdS_4\times\mathbb{CP}^3$ superstring", Mod. Phys. Lett. \textbf{A25} (2010) 1251, \href{http://arxiv.org/abs/0912.1044}{\texttt{arXiv:0912.1044 [hep-th]}}.

\bibitem{Pesando}
I.~Pesando, "A $\kappa$ gauge fixed type IIB superstring action on $AdS_5\times S^5$", JHEP \textbf{9811} (1998) 002,
\href{http://arxiv.org/abs/hep-th/9808020}{\texttt{arXiv:hep-th/9808020}}.

\bibitem{KalloshRahmfeld}
R.~Kallosh and J.~Rahmfeld, "The GS string action on $AdS_5\times S^5$", Phys.\ Lett. \textbf{B443} (1998) 143, \href{http://arxiv.org/abs/hep-th/9808038}{\texttt{arXiv:hep-th/9808038}}.

\bibitem{KT98}
R.~Kallosh and A.A.~Tseytlin, "Simplifying superstring action on $AdS_5\times S^5$", JHEP \textbf{9810} (1998) 016,
\href{http://arxiv.org/abs/hep-th/9808088}{\texttt{arXiv:hep-th/9808088}}.

\bibitem{MT2000}
R.R.~Metsaev and A.A.~Tseytlin, "Superstring action in $AdS_5\times S^5$: $\kappa-$symmetry light cone gauge", Phys.\ Rev. \textbf{D63} (2001) 046002, \href{http://arxiv.org/abs/hep-th/0007036}{\texttt{arXiv:hep-th/0007036}}.\\
R.R.~Metsaev, C.B.~Thorn and A.A.~Tseytlin, "Light-cone
superstring in AdS space-time", Nucl.\ Phys.\ \textbf{B596} (2001)
151, \href{http://arxiv.org/abs/hep-th/0009171}{\texttt{arXiv:hep-th/0009171}}.

\bibitem{BPR}
I.~Bena, J.~Polchinski and R.~Roiban, "Hidden symmetries of the $AdS_5\times S^5$ superstring", Phys.\ Rev. \textbf{ D69} (2004) 046002, \href{http://arxiv.org/abs/hep-th/0305116}{\texttt{arXiv:hep-th/0305116}}.

\bibitem{Minahan}
J.A.~Minahan and K.~Zarembo, "The Bethe-ansatz for $\mathcal N = 4$ super Yang-Mills", JHEP \textbf{0303} (2003) 013,  \href{http://arxiv.org/abs/hep-th/0212208}{\texttt{arXiv:hep-th/0212208}}.

\bibitem{Beisert}
N.~Beisert et. al., "Review of AdS/CFT Integrability, An Overview", \href{http://arxiv.org/abs/1012.3982}{\texttt{arXiv:1012.3982 [hep-th]}}.

\bibitem{Adam}
I.~Adam, A.~Dekel, L.~Mazzucato and Y.~Oz, "Integrability of Type II superstrings on Ramond-Ramond backgrounds in various dimensions", JHEP \textbf{0706} (2007) 085, \href{http://arxiv.org/abs/hep-th/0702083}{\texttt{arXiv:hep-th/0702083}}.

\bibitem{BSZ}
A.~Babichenko, B.~Stefanski and K.~Zarembo, "Integrability and the $AdS_3/CFT_2$ correspondence", JHEP \textbf{1003} (2010) 058, \href{http://arxiv.org/abs/0912.1723}{\texttt{arXiv:0912.1723 [hep-th]}}.

\bibitem{STWZ}
D.~Sorokin, A.~Tseytlin, L.~Wulff and K.~Zarembo, "Superstrings in $AdS_2\times S^2\times T_6$", J. Phys. \textbf{A44} (2011) 275401, \href{http://arxiv.org/abs/1104.1793}{\texttt{arXiv:1104.1793 [hep-th]}}.

\bibitem{0807.0437}
N.~Gromov and P.~Vieira, "The $AdS_4/CFT_3$ algebraic curve", JHEP \textbf{0902} (2009) 040, \href{http://arxiv.org/abs/0807.0437}{\texttt{arXiv:0807.0437 [hep-th]}}.

\bibitem{0807.0777}
N.~Gromov and P.~Vieira, "The all loop $AdS_4/CFT_3$ Bethe ansatz", JHEP \textbf{0901} (2009) 016, \href{http://arxiv.org/abs/0807.0777}{\texttt{arXiv:0807.0777 [hep-th]}}.

\bibitem{Ahn}
C.~Ahn and R.~Nepomechie, "$\mathcal N=6$ Chern-Simons theory S-matrix and all loop Bethe ansatz equations", JHEP \textbf{0809} (2008) 010, \href{http://arxiv.org/abs/0807.1924}{\texttt{arXiv:0807.1924 [hep-th]}}.

\bibitem{SW10}
D.~Sorokin and L.~Wulff, "Evidence for the classical integrability of the complete $AdS_4\times\mathbb{CP}^3$ superstring", JHEP \textbf{1011} (2010) 143, \href{http://arxiv.org/abs/1009.3498}{\texttt{arXiv:1009.3498 [hep-th]}}.

\bibitem{CSW}
A.~Cagnazzo, D.~Sorokin and L.~Wulff, "More on integrable structures of superstrings in $AdS_4\times\mathbb{CP}^3$ and $AdS_2\times S^2\times T_6$ superbackgrounds", JHEP \textbf{1201} (2012) 004, \href{http://arxiv.org/abs/1111.4197}{\texttt{arXiv:1111.4197 [hep-th]}}.

\bibitem{U12}
D.V.~Uvarov, "Kaluza-Klein gauge and minimal integrable extension of $OSp(4|6)/(SO(1,3)\times U(3))$ sigma-model", Int. J. Mod. Phys. \textbf{A27} (2012) 1250118, 
\href{http://arxiv.org/abs/1203.3041}{\texttt{arXiv:1203.3041 [hep-th]}}.

\bibitem{9807115}
G.~Dall'Agata, D.~Fabbri, C.~Fraser, P.~Fre, P.~Termonia and M.~Trigiante, "The $OSp(8|4)$ singleton action from the supermembrane", Nucl. Phys. \textbf{B542} (1999) 157, \href{http://arxiv.org/abs/hep-th/9807115}{\texttt{arXiv:hep-th/9807115}}.

\bibitem{Kallosh2}
R.~Kallosh, "Superconformal actions in Killing gauge", \href{http://arxiv.org/abs/hep-th/9807206}{\texttt{arXiv:hep-th/9807206}}.

\bibitem{PST}
P.~Pasti, D.P.~Sorokin and M.~Tonin, "On gauge-fixed superbrane actions in AdS superbackgrounds", Phys.\ Lett.\  \textbf{B447} (1999) 251, \href{http://arxiv.org/abs/hep-th/9809213}{\texttt{arXiv:hep-th/9809213}}.

\bibitem{U08}
D.V.~Uvarov, "$AdS_4\times\mathbb{CP}^3$ superstring and $D=3$
$\mathcal N=6$ superconformal symmetry", Phys. Rev. \textbf{D79}
(2009) 106007, \href{http://arxiv.org/abs/0811.2813}{\texttt{arXiv:0811.2813 [hep-th]}}.

\bibitem{U-Tomsk}
D.V.~Uvarov, "A note about fermionic equations of $AdS_4\times\mathbb{CP}^3$ superstring", \href{http://arxiv.org/abs/1210.0715}{\texttt{arXiv:1210.0715 [hep-th]}}.

\bibitem{U10}
D.V.~Uvarov, "$D=3$ $\mathcal N=6$ superconformal symmetry of $AdS_4\times\mathbb{CP}^3$ superstring", Class. Quantum Grav. \textbf{28} (2011) 235010, \href{http://arxiv.org/abs/1011.5457}{\texttt{arXiv:1011.5457 [hep-th]}}.



\end{thebibliography}
\end{document}